\documentclass[preprint,showpacs,superscriptaddress,groupedaddress,nofootinbib]{revtex4-1}
\usepackage{graphicx}
\usepackage{bm}       
\usepackage{amssymb}
\usepackage[toc,page]{appendix}
\usepackage{braket}
\newcommand{\del}{\partial}

\usepackage{amsmath}
\renewcommand{\a}{\alpha}
\renewcommand{\b}{\beta}
\begin{document} 
	\title{ Two-dimensional gravity from vanishing metrical dimensions}

\author{Suvikranth Gera}
\email{suvikranthg@iitkgp.ac.in}
\affiliation{Department of Physics, Indian Institute of Technology Kharagpur, Kharagpur-721302, INDIA}

	\author{Sandipan Sengupta}
\email{sandipan@phy.iitkgp.ac.in}
\affiliation{Department of Physics, Indian Institute of Technology Kharagpur, Kharagpur-721302, INDIA}

\begin{abstract}

We obtain a dynamical formulation of two-dimensional gravity from a non-Einsteinian phase in higher dimensions $(D=3+2n)$. The formalism is associated with (at least) one extra dimension of vanishing proper length, thus being inequivalent to either a Kaluza-Klein compactification or the Mann-Ross dimensional reduction defined upon a singular limit. The emergent solutions admit any arbitrary curvature in contrast with Jackiw-Teitelboim constant curvature gravity. We present the static and homogeneous solutions as explicit examples. The effective field equations are shown to remain unaffected by the inclusion of higher Lovelock terms beyond Einstein.
\end{abstract}

	\maketitle 
	\section{Introduction}
	In two spacetime dimensions, the Hilbert-Einstein action is topological, leading to no local gravitational dynamics. At the level of equations of motion, this peculiarity gets reflected through the vanishing of Einstein tensor identically \cite{collas}. This implies that while the spacetime curvature could be arbitrary, the energy-momentum tensor in the sense of Einstein equations must be trivial. 
	This is quite different from the four dimensional case where Einstein gravity emerges as the unique geometric theory, defined by the lowest-order Lovelock term \cite{lovelock,*lovelock1,lanc} depending upon the curvature tensor. 
	
	Here, we confront this problem of defining a unique metric theory in two dimensions from the perspective of gravity in the presence of extra dimensions of vanishing proper length. Such a formulation has only been introduced recently in the context of $D>4$-dimensional gravity theory as an attempt to provide a geometric resolution to the ``dark matter" problem \cite{sengupta}. The framework was developed around the observation that the emergent field content in the presence of invisible (``dark") extra dimensions exhibits  coupling properties quite distinct from ordinary particulate matter and could explain certain special features of the Galactic halo and flat rotation curves. This general framework has also been applied to generate effective dynamics in four dimensions from the otherwise nondynamical Lovelock (Gauss-Bonnet and higher-order) terms \cite{sengupta1}.
	
	Efforts to develop a theory of two-dimensional gravity have a long history  \cite{jackiw,*jackiwt,jackiw1,*jackiw1t,henn,isler,*islert, mann,*mannt1,*mannt2,mann1,*mann1t1,*mann1t2}. Jackiw and Teitelboim \cite{jackiw,*jackiwt} in the early 1980s had proposed  that an appropriate $2D$ analogue of Einstein gravity should be a scalar equation $R+\Lambda=0$, $\Lambda$ being the cosmological constant. The solutions represent constant curvature spacetimes. Within an action principle, this field equation is recovered by introducing a scalar which itself shows up in  a second-order equation \cite{jackiw1,*jackiw1t}. A few years later, Mann and Ross \cite{mann,*mannt1,*mannt2} invoked a dimensional regularization prescription in a classical sense,  reproduce a version of scalar tensor gravity associated with a conserved energy-momentum tensor. 
	Their procedure is based on the singular rescaling of the D dimensional gravitational coupling constant $K_D$, under the assumption that it vanishes as $(D-2)$ in the limit $D\rightarrow 2$. The $D$-dimensional  bimetric action involves two conformally related metrics and leads to an effective action only after the subtraction of a divergent contribution. However, this trick should be viewed as formal  rather than one with a straightforward physical interpretation \cite{sengupta1}. 
	Further, the limit $D\rightarrow2$ essentially attempts to force a connection between a gravity theory with an extra dimensional space of nonzero metrical volume (before the limit) with another where such a subspace has trivial proper size (after the limit).  However, these two gravity theories are strictly inequivalent and could  be (nonanalytically) connected only by a singular diffeomorphism. 
	In retrospect, it is thus not surprising that the Mann-Ross limit invoked to define an effective lower dimensional theory requires a rescaling that is singular and exhibits an action that is divergent without regularization.

Here, we show that within the recent formulation of gravity in the presence of extra dimensions of vanishing proper length, Einstein gravity becomes dynamically nontrivial in two dimensions. 
The emergent theory is shown to be more general than Jackiw- Teitelboim constant curvature gravity and is also inequivalent to Mann-Ross 2D gravity. From the geometric perspective, this theory could be interpreted as being characterized by two diad fields, among which only one is dynamical (associated with a second-order equation). Alternatively, the field content other than the two-metric could be seen to contribute through an effective energy-momentum tensor to the 1+1-dimensional field equations.

Our formalism could also be generalized to include Lovelock densities higher than Einstein by having additional dimensions of vanishing proper length. Remarkably though, such higher-order curvature nonlinearities contribute in a way which precisely reproduces the theory obtained without them the in presence of one and only one vanishing metrical dimension. In other words, the resulting emergent theory is unique with respect to the inclusion of higher Lovelock densities in a higher number of dimensions.

	Let us emphasize that this formulation requires no singular (Mann-Ross) limit such as $D\rightarrow 2$ and no regularization of divergences. Rather than being treated as fictitious, the extra dimensions represent the zero eigendirections of a spacetime with a  noninvertible metric (with one or more zero eigenvalues). Thus, the full spacetime exhibits a subspace whose metrical size vanishes exactly. In general, such spacetimes are known to occur as explicit solutions \cite{tseytlin,kaul,kaul1} to the vacuum field equations within the first-order formalism of gravity theory, which admits invertible as well as noninvertible metric phases \cite{horowitz}.

	In the next section, we elucidate the dimensional reduction of three-dimensional gravity theory where one of the directions has zero proper length, corresponding to a vanishing eigenvalue of the triad. The most general spacetime solutions associated with this noninvertible triad are presented. The resulting emergent 2D gravity theory is discussed, along with a comparison with the earlier formulations in this particular context. The solutions of the emergent field equations for the cases of static and homogeneous spacetimes are presented. Finally, we analyze the critical question of the uniqueness of this theory and demonstrate that the inclusion of  higher-order Lovelock terms  does not affect the emergent field equations. The concluding remarks reflect on the possible relevance of this work in more general contexts.
	
	\section{Three-dimensional action and its reduction}
	
	Let us consider the three-dimensional action with a cosmological constant term:
	\begin{equation*}
		{\cal L}(\hat{e},\hat{w})= \epsilon^{\mu\nu\alpha}\epsilon_{IJK}\left[\xi \hat{e}_\mu^I \hat{R}_{\nu\a}^{\;\;\;JK}(\hat{w})+\frac{\b}{3}\hat{e}_\mu^I\hat{e}_\nu^J\hat{e}_\a^K\right]
	\end{equation*}
	The above reflects a pair of independent variables, e.g. the triad and connection $\hat{e}_\mu^I,~\hat{w}_\mu^{~IJ}$ in three dimensions.  $\xi,~\beta$ are the gravitational coupling and cosmological constant, respectively.
	The associated equations of motion are displayed below:
	\begin{eqnarray}
		\epsilon^{\mu\nu\alpha}\epsilon_{IJK} \hat{D}_\mu(\hat{w}) \hat{e}_\nu^I=0 \label{eom1}\\
		\epsilon^{\mu\nu\alpha}\epsilon_{IJK}\left[\xi \hat{R}_{\mu\nu}^{\;\;\;IJ}(\hat{w}) +\beta \hat{e}^I_{\mu} \hat{e}^J_{\nu}\right]=0\label{eom2}
	\end{eqnarray} 
	Here, we  explore the solution space corresponding to triad fields with one vanishing eigenvalue, which would be assumed to lie along the direction $v$ in the (gauge-invariant) metric: $\hat{g}_{v\mu}=0$. The associated triad in its simplest possible form could be written as
	\begin{eqnarray}
		\hat{e}^I_{\mu}= \begin{bmatrix}
			\hat{e}^i_a\equiv e^i_a &~~ \hat{e}_a^2=0\\
			\hat{e}_v^i=0 &~~ \hat{e}_v^2=0
		\end{bmatrix}  \nonumber
	\end{eqnarray}
The spacetime and internal indices are defined as: $[\mu\equiv (t,x,v)\equiv (a,v)]$ and $[I\equiv (0,1,2)\equiv (i,2)]$.
The diad fields $e_a^i$ with a nonvanishing determinant $e$ could be associated with the emergent two-dimensional spacetime.
	We define the inverse diad fields as $e_i^a$ with $e^a_i e_b^i= \delta^a_b$, $e^a_i e_a^j= \delta^j_i$ and the emergent antisymmetric densities as $\epsilon^{vab}\equiv \epsilon^{ab}$, $\epsilon_{2ij}\equiv \epsilon_{ij}$.
	
	\subsection{Solution to the connection equations:}
	The solution to various components of the connection equations of motion \eqref{eom1} are given below:
	\begin{eqnarray}\label{con-eq}
		\a &=&v, ~(J,K)=(j,2):~~ \epsilon^{ab}\epsilon_{ij}\hat{D}_a \hat{e}_b^i=0= \hat{D}_{\left[a\right.} \hat{e}_{\left.b\right]}^i \implies K_a^{~ij}\equiv \hat{w}_a^{~ij}-\bar{w}_a^{~ij}(e)=0\nonumber		\\
		\a &=&v, ~(J,K)=(j,k):~~\epsilon^{ab}\epsilon_{jk}\hat{D}_a \hat{e}_b^2=0= \hat{w}_{\left[a\right.}^{\;~ 2i} \hat{e}_{\left.b\right]}^i \implies \hat{w}_a^{~2i}=e_{ak} M^{ik}\equiv M_a^i\nonumber\\
		\a &=&b, ~(J,K)=(j,k):~~\epsilon^{ab}\epsilon_{jk}\hat{D}_{\left[a\right.} \hat{e}_{\left.v\right]}^2=0= \hat{w}_{v}^{~2i} \hat{e}_{a}^i \implies \hat{w}_v^{~2i}=0\nonumber\\
		\a &=&b, ~(J,K)=(j,2):~~ \epsilon^{ab}\epsilon_{ij}\hat{D}_{\left[a\right.} \hat{e}_{\left.v\right]}^i=0= \hat{D}_{v} \hat{e}_{a}^i \implies \hat{w}_v^{~ij}=-e_j^a \partial_v e_a^j
	\end{eqnarray}
	where in the first line we have defined $K_a^{~ij}$ as the contortion, and  $\bar{w}_a^{~ij}(e)$ as the torsionless connection completely given by the diads [$\bar{D}_{[a}(\bar{w})e_{b]}^i=0$]. In the second line $M^{ik}=M^{ki}$ is a $2\times 2$ matrix arbitrary up to the triad equations of motion. The last equation implies that the diad determinant $e$ is independent of the third coordinate associated with a null eigenvalue: $e_i^a\partial_v e_a^i=0=\partial_v e$. Since the emergent gauge invariant two-metric $g_{ab}=e_a^i e_{bi}$ is $v$-independent, one must be able to gauge away any apparent $v$-dependence of $e_a^i$. This could be done through the following gauge choice, as evident from the last equation in (\ref{con-eq}):
	\begin{equation}
		\hat{w}_v^{~ij}=0~.
	\end{equation}
	\subsection{Solution to triad equations of motion:}
	Let us consider the remaining field equations \eqref{eom2} here, which are decomposed as below:
	\begin{eqnarray}
		\a&=& a, k=i: ~~\epsilon^{ab}\epsilon_{ij}\hat{R}_{va}^{~~2i}~=~0\implies \partial_v\hat{w}_a^{2i}=0= \partial_vM^{ij}\label{eom3}\\
		\a&=& a, k=2: ~~\epsilon^{ab}\epsilon_{ij}\hat{R}_{va}^{~~ij}~=~0\label{b2}\\
		\a&=& v, k=i:~~ \epsilon^{ab}\epsilon_{ij}\hat{R}_{ab}^{~~2i}~=~0~=~ \bar{D}_{\left[a\right.}M_{\left.b\right]}^i\label{meq}\\
		\a&=& v, k=2:~~\epsilon^{ab}\epsilon_{ij}\left[\xi \hat{R}_{ab}^{~~ij}+\beta e_a^i e_b^j\right]~=~0~=~\epsilon^{ab}\epsilon_{ij}\left[\xi \bar{R}_{ab}^{~~ij}-2\xi M_a^i M_b^j+\beta e_a^i e_b^j\right]\label{mastereq}
	\end{eqnarray}
	In the last two equalities we have used the vanishing of torsion as reflected by eq.(\ref{con-eq}) to replace $\hat{w}_a^{~ij}$ by $\bar{w}_a^{~ij}(e)$. 
	Note that whereas Eq\eqref{eom3} simply reflects the $v$-independence of the field $M_{ab}\equiv M^{ij}e_a^i e_b^j=M_{ba}$, the set(\ref{b2}) is satisfied identically. The only dynamical equation is the last one, which has second order derivatives of the two-metric $g_{ab}$. 
	
	Using the redefined fields above, the equation of motion \eqref{mastereq} reads:
	\begin{equation}\label{Rbar}
		\bar{R}(\bar{w}(e))+\frac{\beta}{\xi}= M^2-M_{a}^i M^{a}_i
		\end{equation} 
		where we have defined the Ricci scalar derived from the torsionless connection $\bar{w}_a^{~ij}(e)$ as $\bar{R}(\bar{w}(e))\equiv e^a_i e^b_j \bar{R}_{ab}^{~~ij}(\bar{w})$ and the trace as $M\equiv e^a_i M_a^i$.
	This, along with Eq.(\ref{meq}), summarizes the main content of the emergent gravity theory on a line, built upon the formalism of extra dimensions of vanishing proper length.
	
	Let us note that the fields $M_{ab}$ may be decomposed in general as:
	\begin{equation}
		M_{ab}= \phi g_{ab}+ S_{ab},\label{Mdecomposition}
	\end{equation}
	where its three components are traded for a scalar $\phi$ and a symmetric traceless field $S_{ab}$ ($g^{ab}S_{ab}=0$). For $S_{ab}=0$,
	Eq \eqref{meq} implies that $\phi$ is constant: $M_{ab}=\lambda g_{ab}$ ($\lambda\equiv const.$). In this case, the equation of motion (\ref{Rbar}) simplifies to:
	\begin{equation}
		\bar{R}+\left[\frac{\beta}{\xi}-2 \lambda^2\right]=0.
	\end{equation}
	Thus, this special case reproduces the Jackiw-Teitelboim 2D gravity equation \cite{jackiw,*jackiwt} upon an identification of $\left[\frac{\beta}{\xi}-2 \lambda^2\right]$ as the (effective) cosmological constant $\bar{\Lambda}$.
	
	Note that it is possible (although not essential) to interpret the field $M_a^i$ as a dual diad that is nondynamical. This would provide a geometric interpretation to the emergent theory above. Based on this, it is possible to set up a bimetric formulation of two-dimensional gravity, which is not explored here any further. The additional field content, however, also admits a nongeometric interpretation in the emergent theory. This is discussed next.
	 
\section{Effective energy-momentum tensor from geometry}
	
	The right-hand side of the field equation (\ref{Rbar}) may (though need not) be interpreted as the effective energy-momentum scalar $\bar{T}\equiv g^{ab}\bar{T}_{ab}\equiv M^{ab}M_{ab}-M^2$, whose origin is purely geometric. 
 This has the following solution for the associated two-tensor:
	\begin{equation}\label{T-eff}
		\bar{T}_{ab} = M_{ac}M^c_{~b}- M M_{ab}+ T_{ab}
	\end{equation} 
	where $T_{ab}$ is any arbitrary symmetric traceless tensor. In principle, the conservation of $\bar{T}_{ab}$ could be imposed consistently as an additional condition, although the theory itself does not require it. 
	
Note that this is in contrast with the original Jackiw-Teitelboim lineal gravity which does not admit a conserved energy-momentum tensor. The Mann-Ross singular limit, while admitting such a tensor, leads to a scalar-tensor gravity where the scalar shows up in a nontrivial second order equation. In our formulation, given the purely geometric two-tensor $\bar{T}_{ab}$ whose definition involves only nonpropagating fields, the geometry of spacetime is determined completely. This is similar in spirit to the Einstein gravitational dynamics in four dimensions where a stress-tensor dictates the curvature of spacetime. 	
	
	Using the general decomposition \eqref{Mdecomposition} we obtain:
	\begin{eqnarray*}
		\bar{T}_{ab}= -2 \phi^2 g_{ab} +S_{ac}S^c_{~b}+ T_{ab} = \overset{(\phi)}{\bar{T}}_{ab}+ \overset{(s)}{\bar{T}}_{ab}+\overset{(t)}{\bar{T}}_{ab}
	\end{eqnarray*}
	In order to unravel the physical properties of this field content, we assume an ideal fluid form for this geometric tensor: $
		\overset{(i)}{\bar{T}}_{ab}= (\bar{\rho}_i+\bar{P}_i)u_a u_b +\bar{P}_i g_{ab} $, $u^a$ being the two-velocity of the fluid and $\bar{\rho}_i,~\bar{P}_i$ being the density and pressure of the $i$-th component. This leads to the following expressions corresponding to the individual components:
		\begin{eqnarray*}
	\bar{P}_S&=&0,~\bar{\rho}_S = -S_{ab}S^{ab}~;\nonumber\\	
\bar{P}_T&=&\bar{\rho}_T~;\nonumber\\
\bar{P}_\phi&=& -\phi^2=-\bar{\rho}_\phi~.
\end{eqnarray*}
Thus, the field multiplet $(S_{ab},~T_{ab},~\phi)$ is composed of an emergent dust, stiff fluid and a spacetime dependent counterpart of the cosmological constant, respectively.

	\section{Effective two dimensional action}
	Our theory may also be reproduced from a purely two dimensional effective action:
	\begin{eqnarray}
		{\cal L}_{eff}(e,w,M,\psi,\Lambda,\bar{\Lambda})~&=&~ \psi \epsilon^{ab}\epsilon_{ij}\left[\xi R_{ab}^{\;\;\;ij}(w)+\beta e_a^i  e_b^j-2\xi M_a^i M_b^j\right]\nonumber\\
		 &+&~2  \epsilon^{ab} \Lambda_i D_a(w) e_b^i~+~2  \epsilon^{ab} \bar{\Lambda}_i D_a(w) M_b^i,
	\end{eqnarray}
	where the fields $\psi,~\Lambda_i,~\bar{\Lambda}_i$ are Lagrange multipliers and $\xi, \beta$ are couplings. The field equations obtained after varying ${\cal L}_{eff}$ with respect to all the independent fields are given by
	\begin{eqnarray*}
	\delta\psi &:&  ~~\epsilon^{ab}\epsilon_{ij}\left[\xi R_{ab}^{\;\;\;ij}(w)~+~\beta e_a^i  e_b^j~-~2 \xi M_a^i M_b^j\right]=0\label{dc}\\
		\delta \Lambda_i &:&~~ \epsilon^{ab}D_a(w) e_b^i=0\label{dl}\\
		\delta \bar{\Lambda}_i &:&~~ \epsilon^{ab}D_a(w) M_b^i=0\label{dm}\\
				\delta w_a^{ij}&:& ~~\epsilon^{ij}\partial_a \psi = \Lambda^{[j} e_a^{i]}+\bar{\Lambda}^{[j} M_a^{i]}\label{eq1}\\
				\delta e_a^i &:& ~~ D_a(w) \Lambda^i=-\beta\epsilon^{ij} e^a_j \psi \label{eq2}\\
				\delta M_a^i &:& ~~ D_a(w) \bar{\Lambda}^i=\xi\epsilon^{ij} M^a_j \psi \label{eq3}
				\end{eqnarray*}
	From the above, we observe that the multipliers obey first order equations among themselves which lead to their solutions. These decouple from the first three equations containing only $e_a^i,w_a^{~ij}$ and $M_a^j$. Upon using the vanishing of torsion as implied by the second equation above, the first and third equations finally become:
	\begin{eqnarray*}
	&&\epsilon^{ab}\epsilon_{ij}\left[\xi \bar{R}_{ab}^{ij}(\bar{w})~-~2 \xi M_a^i M_b^j~+~ \beta e_a^i e_b^j\right]~=~0,\label{eqn1}\\
	&& \epsilon^{ab}\bar{D}_a(\bar{w}) M_b^i~=~0 \label{eqn2}.
	\end{eqnarray*}
	These are precisely the equations of motion \eqref{meq} and \eqref{mastereq} defining the emergent theory. Note that the special case of Jackiw-Teitelboim constant curvature gravity \cite{jackiw1} corresponds to the conditions $M_a^i=0=\bar{\Lambda}$.


\section{Examples}
Here, we  solve a few cases of physical interest, namely, the static and homogeneous cases. These should serve as useful toy models for investigating analogous physics in higher dimensions (e.g. spherical symmetry and cosmological dynamics).

\subsection{Static solutions}
In two dimensions, the most general static two-metric could always be written in the following form using the general coordinate invariance:
\begin{eqnarray*}
ds^2=-f(x)dt^2+\frac{dx^2}{f(x)}
\end{eqnarray*}
We assume that the emergent fields $M_{ij}$ (and hence $M_a^i \equiv M^{ij}e_{aj}$) are static. Using the identities $M^2-M^{a}_i M_{a}^i=2\left[f (M_x^0)^2+M_t^0 M_x^1\right],~\bar{R}(\bar{w}(e))=-f''$, the emergent equations of motion (\ref{meq}) and (\ref{Rbar}) become:
\begin{eqnarray}
&&\del_x M_t^0~=~\frac{f'}{2} M_x^1,~\del_x M_t^1~=~-\frac{f'}{2f} M_t^1,\nonumber\\
&&f (M_{x}^0)^2+M_{t}^0 M_{x}^1~=~-\frac{f''}{2}+\frac{\beta}{2\xi}.
\end{eqnarray}
These three independent equations could be solved for the three independent components of $M_a^i$:
 \begin{eqnarray}
 M_t^1=\frac{C}{\sqrt{f}}=-f M_{x}^0,~M_t^0=\sqrt{\bar{C}+\frac{C^2}{f}+\frac{\beta}{2\xi}f-\frac{f'^2}{4}},~M_x^1=\frac{-\frac{C^2}{f^2}+\frac{\beta}{2\xi}-\frac{f''}{2}}{\sqrt{\bar{C}+\frac{C^2}{f}+\frac{\beta}{2\xi}f-\frac{f'^2}{4}}}
 \end{eqnarray}
 where $C,\bar{C}$ are arbitrary integration constants. Note that the solutions are well defined for $(\bar{C}+\frac{C^2}{f}+\frac{\beta}{2\xi}f-\frac{f'^2}{4})> 0$.
Evidently, given any $f(x)$ defining the static metric, the emergent fields could all be determined.

Next, let us consider the conservation condition:
\begin{eqnarray}\label{conservation}
\nabla_a \bar{T}^{ab}=0=\nabla_a \left[M^{ac} M_{c}^{~b}-MM^{ab}~+~T^{ab}\right]
\end{eqnarray}
These two equations could be solved exactly for the two components of the symmetric traceless tensor $T^{ab}$ (assuming its staticity), using the expressions obtained for $M_{ab}$ earlier:
\begin{eqnarray}\label{T}
\nabla_a \bar{T}^{at}&=&0=\frac{1}{f}\del_x\left[f\bar{T}^{tx}\right]\implies T^{tx}=\frac{k}{f},\nonumber\\
\nabla_a \bar{T}^{ax}&=&0=\sqrt{f}\del_x\left[\frac{\bar{T}^{xx}}{\sqrt{f}}\right]+\frac{ff'}{2}\bar{T}^{tt}\implies 
T^{xx}=-\int dx~f \del_x \left[\frac{M^{xa} M_{a}^{~x}-MM^{xx}}{f}\right]=f^2 T^{tt}\nonumber\\
~
%
\end{eqnarray}
With this, we have the complete solution for all the geometric fields $M_{ab},T_{ab}$ defining the effective energy momentum tensor for any arbitrary spacetime curvature.

The special case $T_{ab}=0$ is of particular interest, which leads to the following solution:
\begin{eqnarray*}
k=0,~f(x)=\left(\bar{k}+\frac{\beta}{2\xi}\right)x^2 + \lambda x +\sigma
\end{eqnarray*}
where $\lambda,\sigma$ are integration constants.
Thus, a trivial $T_{ab}$ corresponds to constant curvature solutions of Jackiw-Teitelboim gravity \cite{jackiw1,*jackiw1t}.

It is straightforward to extend this analysis to the case of two dimensional black holes \cite{mann1,*mann1t1,*mann1t2} [e.g. by replacing $\alpha(x)$ by $\alpha(|x|)$ as an analogue of spherically symmetric solutions], where the curvature singularity should get reflected in the fields $M_{ab}$ through the field equations in our formulation.

\subsection{Homogeneous solutions: Two dimensional cosmology}
Let us now consider a homogeneous geometry in $1+1$ dimensions parametrized by the scale factor $a(t)$: 
\begin{eqnarray*}
ds^2=-dt^2+a^2 (t)dx^2
\end{eqnarray*}
Assuming the fields $M_a^i$ ($M^{ij}$) to be homogeneous, the emergent equations of motion (\ref{meq}) and (\ref{Rbar}) in this case imply:
\begin{eqnarray}
&& \del_t M_x^0+\left[\frac{\dot{a}}{a}\right] M_x^0=0;~~ \del_t M_x^1-\dot{a} M_t^0=0;\nonumber\\
&&\frac{\ddot{a}}{a}+\frac{\beta}{2\xi}-\frac{1}{a^2}\left[(M_{x}^0)^2+a M_{t}^0 M_{x}^1\right]=0.
\end{eqnarray}
These have the following solutions:
 \begin{eqnarray}
 M_{x}^0=\frac{C}{a}&=&-aM_t^1 ,~M_t^0=\frac{\ddot{a}+\frac{\beta }{2\xi}a-\frac{C^2}{a^3}}{\sqrt{\dot{a}^2+\frac{\beta}{2\xi}a^2+\frac{C^2}{a^2}+\bar{C}}},~
 M_x^1=\sqrt{\dot{a}^2+\frac{\beta}{2\xi}a^2+\frac{C^2}{a^2}+\bar{C}}\nonumber\\
~~
 \end{eqnarray}
 where $C,\bar{C}$ are arbitrary integration constants and the solutions are well defined for $(\dot{a}^2+\frac{\beta}{2\xi}a^2+\frac{C^2}{a^2}+\bar{C})> 0$.
 
 Next, we proceed to analyze the consequences of the conservation condition (\ref{conservation}) (assuming homogeneity of the fields $T^{ab}$), whose solution reads:
\begin{eqnarray}\label{T}
T^{tx}(t)=\frac{k}{a^3},~T^{tt}(t)=-\frac{1}{a^2}\int dt~a^2\del_t\left[M^{tc} M_{c}^{~t}-MM^{tt}\right]=a^2 T^{xx}(t)
%
\end{eqnarray}

For the case $T^{ab}=0$, the conservation condition reduces to:
\begin{eqnarray*}
\ddot{a}-\left(\bar{k}-\frac{\beta}{2\xi}\right)a=0
\end{eqnarray*}
where $\bar{k}$ is an integration constant. This admits the following solutions:
\begin{eqnarray*}
a(t)&=&A\cosh{\mu t}+ B \sinh{\mu t}~~\left[\mu^2=\bar{k}-\frac{\beta}{2\xi}>0\right];\\
a(t)&=&C\cos{\omega t}+ D \sin{\omega t}~~\left[\omega^2=\frac{\beta}{2\xi}-\bar{k}>0\right].
\end{eqnarray*} 
Again, this particular case corresponds to spacetimes whose curvature is constant (given by $(\bar{k}-\frac{\beta}{2\xi})$). 

\section{Generalization to higher Lovelock terms: Uniqueness of emergent theory}
In the presence of Lovelock terms higher than the Einstein in a $D\geq 5$-dimensional action, one should expect higher order curvature nonlinearities to appear in the emergent theory, in general. Here, we consider a five dimensional Lovelock theory in order to include the quadratic Gauss-Bonnet term, and explore the resulting emergent theory (in two dimensions) after a dimensional reduction along the lines demonstrated earlier.

The Lagrangian density now reads \cite{sengupta1}\footnote{A different dimensional reduction of this $5D$ action (along with  solutions given by a five-metric with one zero eigenvalue) has recently been considered in ref.\cite{sengupta1} in a formulation of Einstein-Gauss-Bonnet effective theory in four dimensions.}:
\begin{eqnarray}\label{l7}
{\cal L}(\hat{e},\hat{w})&=&\epsilon^{\mu\nu\alpha\beta\gamma} \epsilon_{IJKLM}~\left[\alpha\hat{R}_{\mu\nu}^{~~IJ}(\hat{w})\hat{R}_{\alpha\beta}^{~~KL}(\hat{w})\hat{e}_{\gamma}^{M}
~+~\frac{\chi}{3}\hat{R}_{\mu\nu}^{~~IJ}(\hat{w})\hat{e}_\alpha^K \hat{e}_\beta^L \hat{e}_{\gamma}^{M} ~+~\frac{\beta}{5}\hat{e}_{\mu}^{I}\hat{e}_{\nu}^{J}\hat{e}_{\alpha}^{K}\hat{e}_{\beta}^{L}\hat{e}_{\gamma}^{M}\right]\nonumber\\
~
\end{eqnarray} 
Variation with respect to the five dimensional connection and vielbein fields leads to the following set of equations of motion:
 \begin{eqnarray}
&&\epsilon^{\mu\nu\alpha\beta\gamma} \epsilon_{IJKLM}~\left[\chi\hat{e}_{\alpha}^{I}\hat{e}_{\beta}^{J}~+~2\alpha\hat{R}_{\alpha\beta}^{~~IJ}(\hat{w})\right]\hat{D}_{\mu}(\hat{w})\hat{e}_\nu^K=0,
\label{eom4}\\
&&\epsilon^{\mu\nu\alpha\beta\gamma} \epsilon_{IJKLM}\left[\alpha\hat{R}_{\mu\nu}^{~~IJ}(\hat{w})\hat{R}_{\alpha\beta}^{~~KL}(\hat{w})~+~ \chi\hat{R}_{\mu\nu}^{~~IJ}(\hat{w})\hat{e}_{\alpha}^{K}\hat{e}_{\beta}^{L}
~+~
\beta\hat{e}_{\mu}^{I} \hat{e}_{\nu}^{J}\hat{e}_{\alpha}^{K}\hat{e}_{\beta}^{L}\right]=0.
\label{eom5}
\end{eqnarray}
The full spacetime now has three dimensions of vanishing proper length associated with the three zero eigenvalues of the five dimensional vielbein. Hence we adopt a more general notation following ref.\cite{sengupta1}. The spacetime and internal indices respectively are given by: $\mu\equiv(a,\bar{a})$, $I\equiv (i,\bar{i})$ where $a,i$ are the two dimensional indices and $\bar{a}\equiv (v_1,v_2,v_3),~\bar{i}\equiv (2,3,4)$ are the extra dimensional ones. The only nontrivial components of the degenerate vielbein are $\hat{e}_a^i\equiv e_a^i$, which also denote the emergent diad fields with a nonvanishing determinant $e$: 
\begin{eqnarray}
		\hat{e}^I_{\mu}= \begin{bmatrix}
			\hat{e}^i_a\equiv e^i_a & ~~\hat{e}_a^{\bar{i}}=0\\
			\hat{e}_{\bar{a}}^i=0 & ~~\hat{e}_{\bar{a}}^{\bar{i}}=0
		\end{bmatrix}  \nonumber
	\end{eqnarray}

Let us first find the most general solution to the connection equations (\ref{eom4}). Their decomposition into various components and the corresponding solutions are displayed below:
 \begin{eqnarray}\label{ceq}
\gamma=b,~(L,M)=(i,j)&:&~~\epsilon^{\bar{a}\bar{b}\bar{c}ab} \epsilon_{\bar{i}\bar{j}\bar{k}ij}~\hat{R}_{\bar{a}\bar{b}}^{~~\bar{i}\bar{j}}\hat{D}_{\bar{c}}(\hat{w})\hat{e}_a^{\bar{k}}=0;\nonumber\\
\gamma=\bar{c},~(L,M)=(i,j)&:&~~\epsilon^{\bar{a}\bar{b}\bar{c}ab} \epsilon_{\bar{i}\bar{j}\bar{k}ij}~\hat{R}_{\bar{a}\bar{b}}^{~~\bar{i}\bar{j}}\hat{D}_{a}(\hat{w})\hat{e}_b^{\bar{k}}=0;\nonumber\\
\gamma=b,~(L,M)=(\bar{k},i)&:&~~\epsilon^{\bar{a}\bar{b}\bar{c}ab} \epsilon_{\bar{i}\bar{j}\bar{k}ij}~\hat{R}_{\bar{a}\bar{b}}^{~~\bar{i}\bar{j}}\hat{D}_{\bar{c}}(\hat{w})\hat{e}_a^j=0;\nonumber\\
\gamma=\bar{c},~(L,M)=(\bar{k},i)&:&~~\epsilon^{\bar{a}\bar{b}\bar{c}ab} \epsilon_{\bar{i}\bar{j}\bar{k}ij}~\hat{R}_{\bar{a}\bar{b}}^{~~\bar{i}\bar{j}}\hat{D}_{a}(\hat{w})\hat{e}_b^j=0.
\end{eqnarray}
The remaining components represented by $[\gamma=b,~(L,M)=(\bar{j},\bar{k})]$ and $[\gamma=\bar{b},~(L,M)=(\bar{j},\bar{k})]$ are both satisfied identically upon using these equations above.
Assuming the field-strength components $\hat{R}_{\bar{a}\bar{b}}^{~~\bar{i}\bar{j}}$ as arbitrary, the most general solutions of these equations are obtained as below:
\begin{eqnarray}\label{ceq1}
\hat{D}_{\bar{a}}(\hat{w})e_{b}^{\bar{k}}&=&0\Rightarrow \hat{w}_{\bar{a}}^{~\bar{i}j}=0;\nonumber\\
\hat{D}_{[a}(\hat{w})e_{b]}^{\bar{i}}&=& 0\Rightarrow \hat{w}_a^{~\bar{i}i}=\overset{\bar{(i)}}{{M}^{ik}}e_{ak}~[\overset{\bar{(i)}}{{M}^{ik}}=\overset{\bar{(i)}}{{M}^{ki}}],\nonumber\\
\hat{D}_{[\bar{a}}(\hat{w})e_{b]}^{i}&=& 0\Rightarrow \hat{w}_{\bar{a}}^{~ij}=-e^a_j\del_{\bar{a}} e_a^i,\nonumber\\
\hat{D}_{[a}(\hat{w})e_{b]}^{i}&=&0\Rightarrow K_a^{~ij}\equiv \hat{w}_a^{~ij}-\bar{w}_a^{~ij}(e)=0,
\end{eqnarray}
The third solution above implies that $\hat{w}_{\bar{a}}^{~ij}$ is a pure gauge, which may be fixed to zero using exactly the same argument provided earlier. With this, the emergent diad field is manifestly independent of the extra dimensional coordinates. Note that this set of solutions leads the following field-strength components to vanish:
\begin{eqnarray}
\hat{R}_{\bar{a}\bar{b}}^{~~i\bar{i}}=0=\hat{R}_{\bar{a}\bar{b}}^{~~ij}.
\end{eqnarray} 

Next, we analyze the vielbein equations (\ref{eom5}) and present their general solutions below:
  \begin{eqnarray}\label{eom8}
\gamma=a,~M=i&:&~~\epsilon^{\bar{a}\bar{b}\bar{c}ab} \epsilon_{\bar{i}\bar{j}\bar{k}ij}~\hat{R}_{\bar{a}\bar{b}}^{~~\bar{i}\bar{j}}\hat{R}_{b\bar{c}}^{~~j\bar{k}} =0\implies \hat{R}_{b\bar{c}}^{~~j\bar{k}}=0;\nonumber\\
\gamma=b,~M=\bar{k}&:&~~\epsilon^{\bar{a}\bar{b}\bar{c}ab} \epsilon_{\bar{i}\bar{j}\bar{k}ij}~\hat{R}_{\bar{a}\bar{b}}^{~~\bar{i}\bar{j}}\hat{R}_{a\bar{c}}^{~~ij} =0\implies \hat{R}_{a\bar{c}}^{~~ij}=0;\nonumber\\
\gamma=\bar{c},~M=j&:&~~\epsilon^{\bar{a}\bar{b}\bar{c}ab} \epsilon_{\bar{i}\bar{j}\bar{k}ij}~\hat{R}_{\bar{a}\bar{b}}^{~~\bar{i}\bar{j}}\hat{R}_{ab}^{~~i\bar{k}} =0\implies \hat{R}_{ab}^{~~i\bar{k}} =0;\nonumber\\
\gamma=\bar{c},~M=\bar{k}&:&~~\epsilon^{\bar{a}\bar{b}\bar{c}ab} \epsilon_{\bar{i}\bar{j}\bar{k}ij}~\hat{R}_{\bar{a}\bar{b}}^{~~\bar{i}\bar{j}}\left[\alpha\hat{R}_{ab}^{~~ij}+\chi e_a^i e_b^j\right] =0\implies \epsilon^{ab}\epsilon_{ij}\left[\alpha\hat{R}_{ab}^{~~ij}+\chi e_{a}^i e_{b}^j\right]=0.\nonumber\\
~
\end{eqnarray}  
Note that in the above, we have used the fact that $\hat{R}_{\bar{a}\bar{b}}^{~~\bar{i}\bar{j}}$ are arbitrary.

In the second solution within the set (\ref{ceq1}), we shall first consider the case where exactly one among the three ($\bar{i}\equiv[2,3,4]$) symmetric fields $\overset{\bar{(i)}}{{M}^{ik}}$ is nonvanishing:
\begin{eqnarray}\label{M1}
\overset{(\bar{i})}{{M}^{ik}}=M^{ik}\delta_2^{\bar{i}}
\end{eqnarray}
With this, the various components of first equation among the set (\ref{eom8}) are solved as:
\begin{eqnarray}
\hat{R}_{b\bar{c}}^{~~j2}&=&0 \implies \del_{\bar{c}} M_b^j=0;\nonumber\\
\hat{R}_{b\bar{c}}^{~~j3}&=&0 \implies \hat{w}_{\bar{c}}^{~23}=0;\nonumber\\
\hat{R}_{b\bar{c}}^{~~j4}&=&0 \implies \hat{w}_{\bar{c}}^{~24}=0.
\end{eqnarray}
Thus, the equations of motion naturally force $M_{ij}$, the only emergent field other than the two-metric, to be independent of the extra dimensional coordinates. The above leaves the components $\hat{w}_{\bar{c}}^{~34}$ arbitrary. The second equation implies $\del_{\bar{c}}\hat{w}_a^{~ij}(e)=0$, which simply reconfirms that our gauge choice $\hat{w}_{\bar{a}}^{~ij}=0$ is the correct one. The third equation in this same set, decomposed into its components, leads to:
\begin{eqnarray}\label{R}
\hat{R}_{ab}^{~~i2}&=&0 \implies \bar{D}_{[a}(\bar{w}) M_{b]}^i=0;\nonumber\\
\hat{R}_{ab}^{~~i3}&=&0\implies \hat{w}_{a}^{~23}=0;\nonumber\\
\hat{R}_{ab}^{~~i2}&=&0 \implies \hat{w}_{a}^{~24}=0
\end{eqnarray}
while leaving $\hat{w}_{a}^{~34}$ arbitrary.
Finally, using the vanishing of torsion as obtained in (\ref{ceq1}), 
the last equation in (\ref{eom8}) may be rewritten as:
\begin{eqnarray}
\bar{R}(\bar{w}(e))+\frac{\chi}{\alpha}=M^2-M_{ab}M^{ab}
\end{eqnarray}
where we have defined $M_{ab}=M^{ij}e_{ai} e_{bj}$ as earlier. Note that this, along with the first in the set (\ref{R}) are precisely the emergent equations of motion [equations (\ref{meq}) and (\ref{Rbar})] obtained earlier without the Gauss-Bonnet term, upto an identfication of the respective couplings as $\frac{\chi}{\alpha}\leftrightarrow\frac{\beta}{\xi}$. As for eq. (\ref{M1}), the only other possibilities to consider are when all the three fields $\overset{\bar{(i)}}{{M}^{ik}}$ being either nonvanishing or trivial. The first case leaves these fields as arbitrary leading to no deterministic emergent theory, and should be discarded. The latter case reduces to Jackiw-Teitelboim gravity, which emerges as a special case of our formulation as already elucidated earlier.

To conclude, the addition of higher order Lovelock terms does not affect the emergent theory, whose form remains unique. This feature is remarkable enough, and may be contrasted with a Kaluza-Klein compactification of a higher dimensional theory where the effective dynamics depends upon the number of extra dimensions as well as on the nature of the compactified space.

	\section{Conclusion}
	We have revisited the problem of defining a metric theory of gravity in two spacetime dimensions. Based on the general idea of extra dimensions of vanishing proper length introduced recently \cite{sengupta} and implementing a dimensional reduction, we have obtained a formulation where  spacetime solutions could exhibit any arbitrary curvature. This formalism is more general than the celebrated Jackiw-Teitelboim gravity exhibiting constant curvature solutions.
	
 Our method is inequivalent to a Kaluza-Klein dimensional reduction of a higher-dimensional action, leading to effective scalar-tensor $2D$ gravity in general. In this context, let us also note that the emergent theory is independent of compactification and is not built upon any singular rescaling of gravitational coupling or a subtraction of divergences from the action in the discontinuous limit $D\rightarrow2$. This is in contrast with some of the earlier prescriptions in the literature, invoked to extract nontrivial dynamical effects from higher dimensional actions (e.g. Mann-Ross dimensional regularisation prescription \cite{mann,*mannt1,*mannt2}).
	
	The emergent lineal theory exhibits a nonpropagating field content apart from the two metric. From a geometric viewpoint, this field could be interpreted as a dual diad that is nondynamical. In this sense, our formulation could act as a basis of two-dimensional bimetric gravity. Alternatively, its contribution could be viewed as an effective energy-momentum tensor. As an ideal fluid, its components have the equations of state: $\frac{P}{\rho}=0,\pm 1$. We have solved the effective equations  for the emergent fields for the cases of static and homogeneous geometries. Analogous to four dimensional gravity, specifying this tensor (along with its conservation) completely determines the spacetime geometry. 
	
	We have explicitly demonstrated the uniqueness of the emergent theory. The inclusion of higher Lovelock terms (Gauss-Bonnet and so on) in presence of more than one ``dark" dimension has no effect on the general form of the emergent $2D$ theory.  
		The fact that only a single  extra dimension of vanishing proper length is  relevant, leading to contributions from only a finite number of Lovelock densities to the emergent field equations, appears to be a generic feature in this dimensional reduction formalism associated with non-invertible vielbein fields\footnote{The case of a four-dimensional degenerate tetrad with two zero eigenvalues, considered in Ref \cite{kaul1}, does not lead to a deterministic emergent theory of 2D gravity as here.}. The recent formulation of a four dimensional Einstein-Gauss-Bonnet effective theory \cite{sengupta1} based on a non-Einsteinian phase supports this observation, even though the original motive to consider this case is different from here.
	
	It is well known that the quantization of lower dimensional gravity does provide important insights \cite{henn,jacobson}, particularly in view of the unresolved issues in four dimensional quantum gravity. One wonders if our formulation here could add to the general wisdom upon a canonical or Wheeler-Dewitt quantization. That might be worthwhile, given the genericity of its solution space as compared to Jackiw-Teitelboim gravity and its essential resemblance to classical gravitational dynamics in four dimensions. Further, it seems plausible that the quantum counterpart of our formulation, either from the $D>2$ or the emergent perspective, could be connected to quantum gravity states that have support only along a single dimension. Explicit examples are the (loop) states defined on  Wilson lines which define a 1+1-dimensional space-time embedded within a noninvertible metric \cite{jacobson}.

\begin{acknowledgments}
It is a pleasure to acknowledge the comments by   A. Virmani  on the manuscript and general discussions with  A.Laddha. S.G. is supported by the Inspire doctoral fellowship funded by SERB, DST.

\end{acknowledgments}
	
	\bibliographystyle{apsrev4-1}
\bibliography{egb_2d}

\end{document}